\documentclass{aa}  
\usepackage{graphicx}
\usepackage{txfonts}

\usepackage{natbib}
\usepackage{amsmath} 
\usepackage{amssymb}
\usepackage{mathtools}
\usepackage{microtype}
\usepackage[dvipsnames]{xcolor}
\definecolor{CiteRed}{RGB}{110, 0, 0}
\usepackage{multirow}
\usepackage{array}
\usepackage{makecell}
\usepackage{siunitx}
\usepackage{xcolor}
\usepackage{orcidlink}
\usepackage{graphicx}
\usepackage[caption=false, position=top]{subfig}
\usepackage{verbatim}
\graphicspath{ {img/} {plot/} {fig/} }
\usepackage{hyperref}
\hypersetup{
    breaklinks,
    colorlinks,
    citecolor=CiteRed,  
    linkcolor=NavyBlue, 
    urlcolor=NavyBlue,  
    linktoc=page,
    pdftitle={A new quasar strongly-lensed candidate by the galaxy cluster WHJ0400--27 with a $18''$ image-separation},
    pdfkeywords={Quasars, Strong gravitational lensing, Cosmology, Astrophysics},
    pdfauthor={Lorenzo Bazzanini},
    pdfcreator={\LaTeX}
}

\begin{document} 

   \title{A new quasar strongly-lensed candidate by the galaxy cluster WHJ0400--27 with a $18''$ image-separation}
    \author{L.~Bazzanini~\inst{1,2}\fnmsep\thanks{\texttt{bzzlnz[at]unife[dot]it}}\orcidlink{0000-0003-0727-0137}\and
        G.~Angora~\inst{3,1}\orcidlink{0000-0002-0316-6562}\and
        M.~Scialpi~\inst{7,8,13}\orcidlink{0009-0006-5100-4986}\and
        G.~Di~Rosa~\inst{1}\orcidlink{0009-0001-9416-0923}\and        P.~Bergamini~\inst{4,2}\orcidlink{0000-0003-1383-9414}\and
        P.~Rosati~\inst{1,2}\orcidlink{0000-0002-6813-0632}\and
        M.~Lombardi~\inst{4}\orcidlink{0000-0002-3336-4965}\and
        D.~Abriola~\inst{4}\orcidlink{0009-0005-4230-3266}\and
        A.~Acebron~\inst{6}\orcidlink{0000-0003-3108-9039}\and
        M.~D'Addona~\inst{12,3}\orcidlink{0000-0003-3445-0483}\and
        G.~Granata~\inst{1,4}\orcidlink{0000-0002-9512-3788}\and
        C.~Grillo~\inst{4,5}\orcidlink{0000-0002-5926-7143}\and
        F.~Mannucci~\inst{8}\orcidlink{0000-0002-4803-2381}\and
        M.~Maturi~\inst{9,10}\orcidlink{0000-0002-3517-2422}\and
        M.~Meneghetti~\inst{2}\orcidlink{0000-0003-1225-7084}\and
        A.~Mercurio~\inst{12,3,14}\orcidlink{0000-0001-9261-7849}\and
        M.~Radovich~\inst{11}\orcidlink{0000-0002-3585-866X}
        }

   \institute{
   Department of Physics and Earth Science, University of Ferrara, via Saragat 1, I--44122, Ferrara, Italy\label{unife}
   \and 
   INAF -- OAS, Osservatorio di Astrofisica e Scienza dello Spazio di Bologna, via Gobetti 93/3, I-40129 Bologna, Italy\label{inafbo}
   \and
   INAF -- Osservatorio Astronomico di Capodimonte, Salita Moiariello 16, I-80131 Napoli, Italy\label{inafna}
   \and
   Dipartimento di Fisica, Universit\`a  degli Studi di Milano, via Celoria 16, I-20133 Milano, Italy\label{unimi}
   \and
   INAF -- IASF Milano, via A. Corti 12, I-20133 Milano, Italy\label{inafmi}
   \and
   Departamento de Física Moderna, Instituto de Fisica de Cantabria, Avd.~Los Castros 48, 39005 Santander, Spain\label{cantabria}
   \and
   Università di Firenze, Dipartimento di Fisica e Astronomia, via G. Sansone 1, 50019 Sesto F.no, Firenze, Italy\label{unifi}
    \and
   INAF -- Osservatorio Astrofisico di Arcetri, Via Largo E. Fermi 5, 50125 Firenze, Italy\label{inaffi}
   \and
   Zentrum f\"ur Astronomie, Universit\"at Heidelberg, Philosophenweg 12, D-69120 Heidelberg, Germany\label{uniheidelberg} 
   \and
   ITP, Universit\"at Heidelberg, Philosophenweg 16, D-69120 Heidelberg, Germany\label{itpheid}
   \and
   INAF -- Osservatorio Astronomico di Padova, vicolo dell'Osservatorio 5, I-35122 Padova, Italy\label{inafpd}
   \and
    Università di Salerno, Dipartimento di Fisica "E.R. Caianiello", Via Giovanni Paolo II 132, I-84084 Fisciano (SA), Italy\label{unisa}
    \and
    University of Trento, Via Sommarive 14, I-38123 Trento, Italy\label{unitn}  
    \and 
    INFN -- Gruppo Collegato di Salerno -- Sezione di Napoli, Dipartimento di Fisica ``E.R. Caianiello'', Università di Salerno, via Giovanni Paolo II, 132 - I-84084 Fisciano (SA), Italy\label{infnsa} 
   }
   
   \date{Received xxx; accepted xxx}

   \abstract
   {Time-delay cosmography (TDC) using multiply-lensed quasars (QSOs) by galaxies has recently emerged as an independent and competitive tool to measure the value of the Hubble constant. Lens galaxy clusters hosting multiply-imaged QSOs, when coupled with an accurate and precise knowledge of their total mass distribution, are equally powerful cosmological probes. However, less than ten such systems have been identified to date.}
   {Our study aims to expand the limited sample of cluster-lensed QSO systems by identifying new candidates within rich galaxy clusters.}
   {Starting from a sample of $\sim 10^5$ galaxy cluster candidates~\citep{wen2022}, built from Dark Energy Survey and Wide-field Infrared Survey Explorer imaging data, and a highly-pure catalogue of over one million QSOs, based on {\it Gaia} DR3 data, we cross-correlate them to identify candidate lensed QSOs near the core of massive galaxy clusters.}
   {Our search yielded 3 lensed double candidates over an area of $\approx 5000$ sq.~degree. In this work, we focus on the best candidate consisting of a double QSO with {\it Gaia}-based redshift of 1.35, projected behind a moderately rich cluster (WHJ0400-27) at $z_{\rm phot}=0.65$. Based on a first spectroscopic follow-up study, we confirm the two QSOs at $z=1.345$, with indistinguishable spectra, and a brightest cluster galaxy at $z=0.626$. These observations seem to support the strong lensing nature of this system, although some tension emerges when the cluster mass from a preliminary lens model is compared with that from other mass proxies. We also discuss the possibility that such system is a rare physical association of two distinct QSOs with a projected physical distance of $\approx 150$ kpc. If further spectroscopic observations confirm its lensing nature, such a rare lens system would exhibit one of the largest image separations observed to date ($\Delta\vartheta=17.8''$), opening interesting TDC applications.}
   {} 
  

   \maketitle



\section{Introduction}
\label{sec::intro}

The Hubble constant ($H_0$) stands as a pivotal cosmological parameter defining the dimensions, age, and expansion rate of the Universe. The increased precision in its determination has revealed significant discrepancies between estimates derived from observations in the local Universe and those derived from early Universe probes~\citep{verde2019, moresco2022}. Therefore, independent and complementary techniques for measuring the value of $H_0$ become crucial to assess this tension, which may hint at the presence of new physics beyond the standard cosmological model.

\citet{refsdal1964} predicted that time delays between multiple images of strongly-lensed supernovae (SNe) or variable sources such as quasars (QSOs) could offer a novel method to measure the value of $H_0$. Since then, time-delay cosmography (TDC) has become a mature technique, providing competitive estimates of this parameter. In particular, QSOs strongly lensed by galaxies have been exploited by the H0LiCOW program~\citep{suyu2017} to measure the value of $H_0$ with a $2.4\%$ precision from the joint analysis of six gravitationally lensed quasars~\citep{wong2020}, each system providing a measurement with precisions ranging from 3.6--9.3\%. 

On the other hand, using time-varying sources strongly lensed by galaxy clusters is a complementary technique that has remained largely unexploited to date, although it has recently proven its sheer potential. SN `Refsdal', discovered by~\citet{kelly2015} to be strongly lensed by the galaxy cluster MACS J1149.5+2223, led~\citet{grillo2024} to infer the value of $H_0$ with a 6\% (statistical plus systematic) uncertainty using a full strong lensing (SL) analysis, which included the measured time delays between the SN multiple images and the associated uncertainties (see also~\citealt{grillo2018, grillo2020}). This study also demonstrated that time delays in lens galaxy clusters are a valuable and complementary tool to measure both $H_0$ and the geometry ($\Omega_{\rm m},\, \Omega_{\mathrm{DE}}, \, w$) of the Universe, by exploiting the observed positions of other multiply-lensed sources at different redshifts. In addition, these allow for a significant reduction of the mass-sheet degeneracy, which plagues TDC with galaxy-scale lens systems~\citep{birrer2016, grillo2020, moresco2022}.

As estimated by~\citet{treu2022}, a sample of $\sim40$ lensed QSO or SNe is expected to yield $H_0$ measurement with $\sim\! 1$\% precision, which can be extremely relevant to assess the $H_0$ tension. However, less than ten QSOs strongly-lensed by galaxy clusters have been discovered to date~\citep{inada2003, inada2006, dahle2013, shu2018, shu2019, martinez2023, napier2023}. Expanding the sample is therefore of paramount importance and can guide the search for such rare systems in upcoming wide-field surveys, such as \emph{Euclid}~\citep{euclid2019} and Rubin-LSST~\citep{LSST2019}. Other studies~\citep{dutta2024} use strongly-lensed QSOs by galaxy clusters to probe the circumgalactic medium, investigating the distribution and coherence of metal-enriched gaseous structures.

As part of a search for gravitationally lensed {\it Gaia} QSOs by galaxy clusters, we present the discovery of a pair of QSOs ($z_{Gaia}\simeq1.35$), possibly lensed by a relatively rich Dark Energy Survey (DES) galaxy cluster at $z\simeq0.63$. If confirmed, the image separation of $\sim18''$ would make it one of the largest separation systems among lensed QSOs known to date. We also consider the possibility for this system to be a dual AGN~\citep[e.g.][]{hennawi2006, mannucci2023}, often considered as supermassive black-hole merger candidates, which would result in a very rare system, given the remarkable similarities of the two QSO spectra. The data used in our study and the implemented search algorithm are detailed in Sec.~\ref{sec::search}. In Sec.~\ref{sec::results}, we present the follow-up ground-based spectroscopy which makes this candidate a likely gravitationally lensed QSO, along with a preliminary SL model of the lens galaxy cluster. In Sec.~\ref{sec::discussion}, we discuss the challenges posed by the SL interpretation. In Sec.~\ref{sec::conclusions}, we summarise our findings for and against the strong lensing scenario.



Throughout the paper, we assume a flat $\Lambda\mathrm{CDM}$ cosmology with $\Omega_\Lambda = 0.7$, $\Omega_{\rm m} = 0.3$, and $H_0 = 70\,\mathrm{km}\,\mathrm{s}^{-1}\,\mathrm{Mpc}^{-1}$. Magnitudes are reported in the AB system, unless otherwise stated.


\section{Data \& Method}
\label{sec::search}
We have conducted a search for strongly-lensed QSOs by cross-matching a sample of galaxy cluster candidates derived from DES imaging data~\citep{wen2022} with a highly pure {\it Gaia} DR3 QSO catalogue~\citep{gaiadr3qso}.

To identify galaxy clusters, \citet{wen2022} combined the DES optical data~\citep{des2016}, covering $\sim5000\,\mathrm{deg}^2$ in the southern sky, with the mid-infrared data from the WISE (Wide-field Infrared Survey Explorer, \citealt{wise2010, lang2014, schlafly2019}) all-sky survey. The final catalogue contains $1.5\times 10^5$ galaxy cluster candidates, with photometric redshift up to $z\simeq1.5$. Clusters were selected based on over-densities of galaxy stellar mass within a given photometric redshift slice, using a nearest-neighbour algorithm.

The {\it Gaia} DR3 QSO catalogue~\citep{gaiadr3qso} contains a total of $6.6 \times 10^6$  candidates, classified using parallax, proper motion, $BP$/$RP$ spectra, brightness, and/or variability. For our analysis we used a highly-pure sub-sample, consisting of $1.7$ million objects with spectro-photometric redshifts, characterised by a magnitude $G \lesssim 20.5$, with a purity estimated by the {\it Gaia} collaboration of $95\%$. The redshifts of the QSOs are computed from {\it Gaia}’s low-resolution $BP$/$RP$ spectra, using a $\chi^2$ approach through comparison to a Sloan Digital Sky Survey (SDSS) composite spectrum.

We then proceeded to cross-match the galaxy cluster and QSO catalogues, searching for clusters with:
\begin{enumerate}
    \item at least two QSOs within $1'$ from the cluster centre, that are behind the cluster, namely ~$(z_{\rm QSO}-z_{\rm CL})/\sigma_z^{\mathrm{QSO-CL}}>2$;
    \item QSO separation $< 30''$;
    \item QSOs with consistent redshifts, namely\\
    $|z_{\rm QSO-A} - z_{\rm QSO-B}|/ \sigma_z^{\mathrm{QSOs}}<1$;
\end{enumerate}
where $\sigma_z$ is the propagated uncertainty.

The search resulted in a total of $3$ candidate lensed QSO pairs over $\approx\! 5000\, {\rm deg^2}$, demonstrating the extraordinary rarity of these systems. Interestingly, such a rate of $\sim 1$ bright multiply lensed QSO per $1000\,{\rm deg^2}$ is consistent with the number of QSOs with separations $>10''$ found in the SDSS~\citep{inada2003}.


\section{Results}
\label{sec::results}
Based on these findings, in this Section we describe the properties of the most likely lens system and the follow-up spectroscopic study of the two QSOs, carried out with two different spectrographs. Finally, we present a preliminary SL model and the search for the third model-predicted QSO counter-image.

\subsection{Lensed QSO candidate}
\begin{figure*}
    \centering
    \includegraphics[width = 0.7\textwidth]{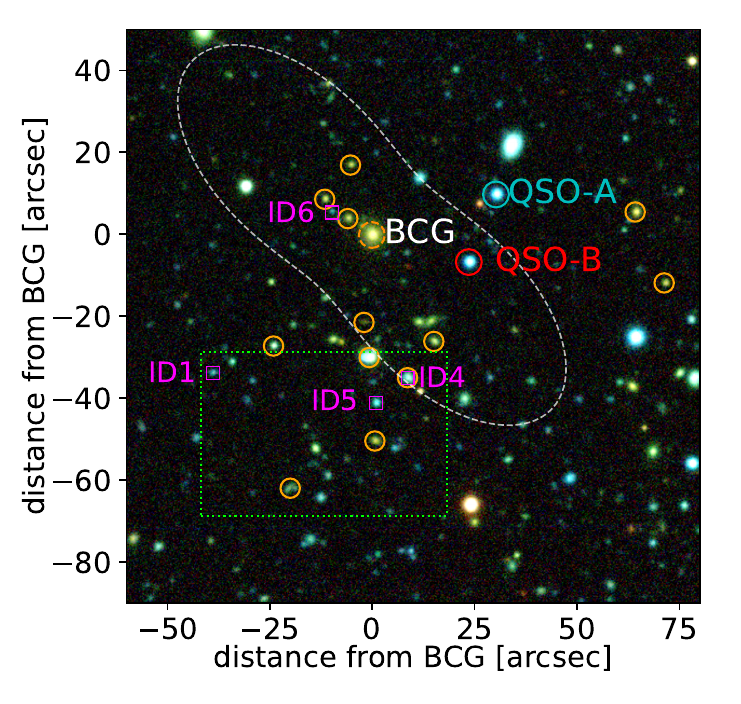}
    \caption{Ground-based $140'' \times 140''$ DES $g$, $r$, $y+z$ image of the galaxy cluster WHJ0400--27, with coordinates centred on the BCG ($z_{\rm cl,spec}=0.626$, $(\mathrm{R.A., Dec.})_{\mathrm{J2000}} = (60.04888, -27.11959)\,\mathrm{deg}$). The cyan and red circles indicate the QSO pair ($z_{\rm spec}=1.345$). The candidates for the third QSO image, with colours consistent with those of the QSOs, are enclosed by magenta squares (ID1-4-5-6). The green dotted rectangle represents the approximate region where the lens model predicts the third image position. Photometric cluster members are marked with orange circles. The white dashed line is the critical line predicted by the best-fit lens model at the QSO redshift. 
    }
    \label{fig::cluster}
\end{figure*}

The most likely lens system shows two candidate QSO multiple images with a  redshift $z_{\mathrm{QSO,} Gaia}\simeq1.35$, lensed by the DES galaxy cluster WHJ040011.7--270711 (hereafter WHJ0400--27) at $z_{\rm phot, DES} = 0.65$, shown in Fig.~\ref{fig::cluster}. The physical properties of the two QSOs, hereafter labelled QSO-A and QSO-B, are presented in Table~\ref{table::coordinates}~\citep{gaiadr3qso}. This system will be referred to as QSO~0400–27. The two putative multiple images are separated by $\Delta\vartheta = 17.8''$, at a projected distance of $31.5''$ and $24.5''$ from the brightest cluster galaxy (BCG) of WHJ0400--27 ($(\mathrm{R.A., Dec.})_{\mathrm{J2000}} = (60.04888, -27.11959)\,\mathrm{deg}$). \cite{wen2022} also provide a catalogue of 11 photometrically identified cluster members associated to this cluster, which they obtained directly from the cluster-finding algorithm. 

By comparing our system with the list provided by~\citet[see their Table~3]{martinez2023}, if confirmed, QSO~0400--27 would be the fifth largest-separation lensed QSO known to date, considering also the most recent one found by~\citet{napier2023}.


\begin{table*}
\caption{Properties of the QSO multiple images based on {\it Gaia} DR3.} 
\label{table::coordinates} 
\centering
    \begin{tabular}{c c c c c c c c c}
    \hline\hline
    Object  & R.A.$^{\dagger}$ & Dec.$^{\dagger}$ & $z_{\it Gaia}^{\,\ast}$  & Mean $G^{\,\ddagger}$ &  Mean $BP^{\,\ddagger}$ &  Mean $RP^{\,\ddagger}$ & $P_{\mathrm{QSO}}^{\,\aleph}$ & $\Delta\theta^{\,\star}$\\
    \hline
    QSO-A   & $60.039429$      & $-27.116843$     & $1.351\pm0.011$          & $19.812\pm0.004$      & $20.01\pm0.04$          & $19.52\pm0.05$          & $100\%$      & $31.5''$\\ 
    QSO-B   & $60.041497$      & $-27.121429$     & $1.35\pm0.02$            & $20.227\pm0.005$      & $20.41\pm0.06$          & $19.82\pm0.07$          & $99.8\%$     & $24.5''$\\
    \hline\hline 
    \end{tabular}
    \tablefoot{Data from~\citet{gaiadr3qso}. 
    \begin{list}{}{}
       \item [$\dagger$] Right ascension and Declination in degrees. All celestial coordinates are given in J2000.
       \item [$\ast$] {\it Gaia} spectro-photometric redshifts.
       \item [$\ddagger$] $G$, $BP$, $RP$ {\it Gaia} magnitudes.
       \item [$\aleph$] {\it Gaia} QSO classification probability.
       \item [$\star$] Angular separation of the QSO from the BCG.
    \end{list}
    }
\end{table*}

\subsection{Spectroscopic confirmation of the QSO multiple images}

To confirm the nature of this double QSO system, we obtained the first long-slit spectra of the QSOs A and B with the Alhambra Faint Object Spectrograph and Camera (ALFOSC, \citealt{djupvik2010}), mounted at the \SI{2.56}{m} Nordic Optical Telescope (NOT) at Roque de los Muchachos Observatory, La Palma, Spain, and with the ESO Faint Object Spectrograph and Camera v2 (EFOSC2, \citealt{buzzoni1984}), mounted at the \SI{3.58}{m} New Technology Telescope (NTT) in Chile. 

QSO-A and -B were spectroscopically observed on November 15, 2023, with the ALFOSC/NOT instrument (program ID: 68-404, PI: Bazzanini). The total integration time was \SI{1}{h} (on-target), using a $1''$--wide longslit, dispersed by the grism \#3 (covering a wavelength range of $3200$--$7070$ \AA), with a resolution of $R \simeq 350$, dispersion of $2.3\,\mathrm{\AA}\,\mathrm{pixel}^{-1}$. The average seeing during the observation was $0.6''$, with an average air mass of 2.1. 

Both QSOs were also observed on February 12, 2024 using the EFOSC2/NTT (program ID: 112.25CT, PI: Mannucci) to obtain a higher signal-to-noise ratio (S/N) and a longer wavelength coverage. The total integration time was \SI{45}{min} (on-target), using a $1''$--wide longslit, dispersed by the \#4 grism (covering a wavelength range of $4085$--$7520$ \AA), with a resolution of $R \simeq 460$, 
dispersion of 1.68 \AA\,$\mathrm{pixel}^{-1}$, an average seeing during the observation of $0.9''$, and an average air mass of 1.1. 

We reduced all our data using the \texttt{PypeIt} data reduction pipeline~\citep{prochaska20}, a Python package for semi-automated reduction of astronomical spectroscopic data. We then flux calibrated the NTT data using the standard star GD108 within the same pipeline. Note that flux calibration of the NOT spectra was not possible due to the lack of spectroscopic standards observations with the adopted grism.

\begin{figure}
    \subfloat[NOT/ALFOSC \SI{60}{min} (on-target).]{%
      \includegraphics[clip,width=0.48\textwidth]{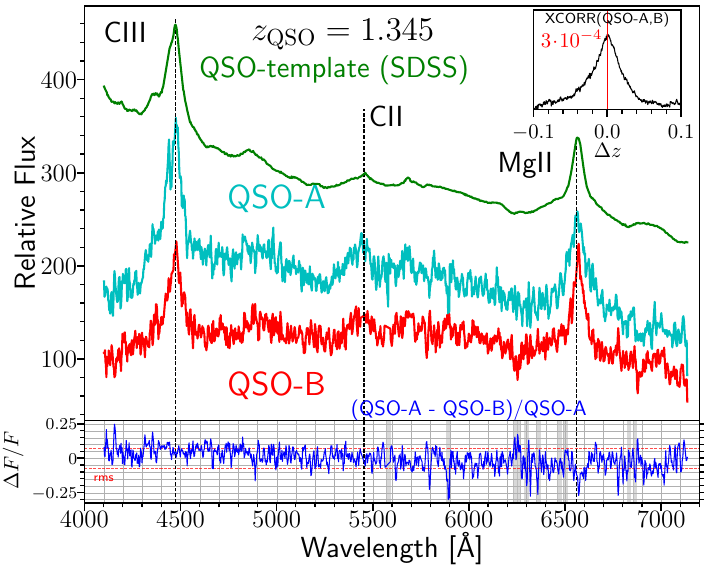}%
    }%
    
    \subfloat[NTT/EFOSC2 \SI{45}{min} (on-target).]{%
      \includegraphics[clip,width=0.48\textwidth]{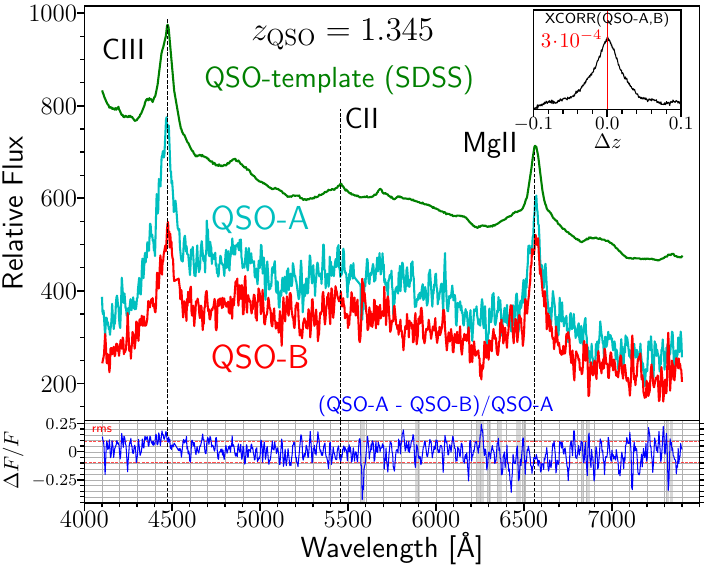}%
    }%
    
    \caption{Spectra of the two candidate lensed QSOs (A and B, resp.~cyan and red), from NOT/ALFOSC (\emph{top}) and NTT/EFOSC2 (\emph{bottom}). The green curve is a SDSS QSO spectrum template. The vertical dotted lines ($4475.97\,\AA$, $5454.47\,\AA$, and $6563.07\,\AA$) represent the positions of emission lines of the corresponding ions redshifted to $z_{\rm QSO, spec}=1.345$ of CIII] ($1908.73\,\AA$), CII] ($2326\,\AA$), and MgII ($2798.75\,\AA$), respectively. The two top-right insets show the cross-correlations of QSO-A and -B spectra. 
    The bottom panel in each plot (blue line) shows the mean subtracted relative flux difference of QSO-A w.r.t~QSO-B, the red dashed lines representing its r.m.s.. The gray bands highlight spectral regions corresponding to prominent sky emission lines, corresponding to significant residuals in the sky subtraction. 
    }
    \label{fig::spectra}
\end{figure}

The reduced QSOs spectra from both the NOT and NTT are shown in Fig.~\ref{fig::spectra} top and bottom, respectively. 
The signal-to-noise per pixel ranges from 5 to 8 (depending on wavelength), slightly higher in the NTT data. 
Both these observations suggest that the two QSO spectra are identical, with no measurable velocity shift, with the same continuum shape, dust extinction, emission line shape, and line flux ratios. 
Indeed, a cross-correlation between the spectrum of each of the two QSOs and a SDSS QSO spectrum template extracted with the PANDORA EZ package~\citep[Easy Z]{garilli2010} yields the same redshift value within $\delta z_{\rm NOT} = 2 \times 10^{-3}$, $\delta z_{\rm NTT} = 3 \times 10^{-4}$, resulting in $z_{\rm QSO, spec}^{\rm NOT}=z_{\rm QSO, spec}^{\rm NTT}=1.345$, confirming the values estimated in the {\it Gaia} catalogue.  
Moreover, the cross-correlation between the two observed QSO spectra (shown in the two top insets in Fig.~\ref{fig::spectra}) peaks at $\Delta z_{\rm NOT} = (3 \pm 8) \times 10^{-4}$, $\Delta z_{\rm NTT} = (3\pm2) \times 10^{-4}$. The errors on the cross-correlation peak have been estimated with a bootstrap procedure, which utilized 100 realisations of the QSO spectra by sampling the error spectrum given by the data reduction pipeline. These values of $\Delta z$, which correspond to a rest-frame velocity difference of $\approx40$ km/s, are therefore fully consistent with zero. This implies no detectable velocity shift between the two spectra, supporting the SL nature of the event. 

The bottom panel of each plot in Fig.~\ref{fig::spectra} shows the mean subtracted relative flux difference of the two QSOs. This shows an r.m.s.~of $7\%$ ($9\%$) for the NOT (NTT) case, supporting the conclusion that the two QSOs exhibit highly similar spectral properties. We note that differential flux calibration errors generally lead to longer wavelengths residuals, as particularly evident in the NOT spectra. 

The inspection of the NOT spectra in Fig.~\ref{fig::spectra}a shows some steepening on the blue side for QSO-A, which is however not present in the NTT spectra. We possibly attribute this effect to differential atmospheric refraction~\citep{filippenko1982}, which we have verified affects the NOT observations in a stronger fashion, given the large air mass and the slit orientation which is far from the parallactic angle. If the QSOs are not equally centred on the slit, this may lead to differential flux losses on the blue side of the spectra. 

A close inspection of the NTT spectra in Fig.~\ref{fig::spectra}b also indicates that the CIII and MgII emission peaks are of comparable strength for QSO-B, whereas they appear to have a different strength for QSO-A ($\approx15\%$ difference), assuming an optimal flux calibration across the entire wavelength range. Should this discrepancy be confirmed by improved spectral observations, in the SL scenario this could be the result of differential levels of dust extinction and/or chromatic microlensing effects (e.g.~\citealt{sluse2012}).

\begin{figure}
    \centering
    \includegraphics[width = 0.48\textwidth]{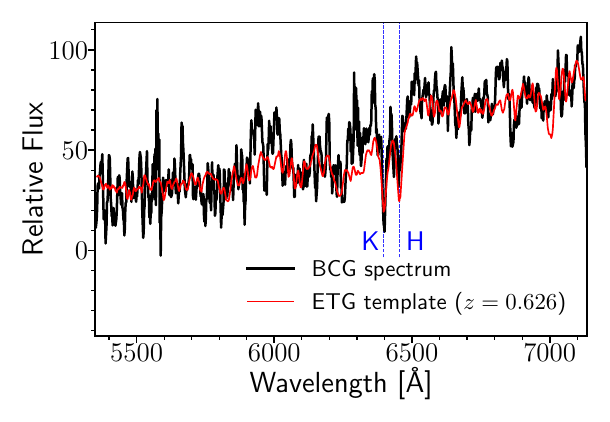}
    \caption{NOT/ALFOSC \SI{60}{min} (on-target) spectrum of the WHJ0400--27 BCG (black line). The red line is a redshifted early-type galaxy (ETG) spectrum template, which provides a significant cross-correlation peak at $z=0.626$. The vertical blue dashed lines represent the positions of the redshifted $\mathrm{CaII}$ K and H absorption lines ($3933.7\,\AA, 3968.5\,\AA$).
    }
    \label{fig::spectrabcg}
\end{figure}

An additional NOT spectrum taken on December 2, 2023 (program ID: 68-404, PI: Bazzanini) also confirmed the DES photometric redshift of the BCG, yielding $z_{\rm CL,spec}=0.626$ (see Fig.~\ref{fig::spectrabcg}). The total integration time was \SI{1}{h} (on-target), using a $1.8''$--wide longslit, dispersed by the grism \#7 (covering a wavelength range of $3650$--$7110$ \AA), with a resolution of $R \simeq 650$, dispersion of $1.7\,\mathrm{\AA}\,\mathrm{pixel}^{-1}$, and an average seeing during the observation of $0.9''$. 

\subsection{Strong lensing modelling}

Assuming that the two QSOs are multiple images of the same background source, the SL model is inevitably poorly constrained, owing to the lack of additional multiple images and spectroscopic information of cluster member galaxies (with the exception of the BCG). The configuration of the double QSO suggests the existence of a third, less magnified, multiple image to the south of the BCG. To search for such an image, we built a preliminary SL model by using a single cluster-scale halo with an elliptical total mass distribution centred on the BCG. To this aim, we used a newly developed lens modelling software, \texttt{Gravity.jl}~\citep{lombardi2024}, and independently validated the results with the public software \texttt{lenstool}~\citep{kneib1996, jullo2007, jullo2009}. 
For the cluster halo, we specifically adopted a non-singular isothermal ellipsoid with velocity dispersion, core radius, axis-ratio and position angle as free parameters; large flat priors were assumed on these parameters, and no priors were adopted on the position of the third image. The range of the flat priors on the free parameters used during the Monte-Carlo Markov Chain (MCMC) optimization are:%
\begin{itemize}%
    \item velocity dispersion: $\sigma \in [100, 3000]\,\mathrm{km/s}$;
    \item core radius: $s \in [10^{-4}, 50]$ arcsec;
    \item axis-ratio: $q \in [0.5, 1]$;
    \item position angle (clockwise from North): $\theta \in [-\pi/2, \pi/2]$.
\end{itemize}%
We used a $0.1''$ positional uncertainty for QSO-A and B. 

The model optimization yielded the following formal values for the median and the $1\sigma$ errors: $\sigma = 2327^{+174}_{-98}$ km/s, $s = 28^{+10}_{-6}$ arcsec, $q = 0.6^{+0.1}_{-0.1}$, $\theta = -0.8^{+0.3}_{-0.2}\,\mathrm{rad}$. 
The optimized model predicts a magnitude difference $\Delta G_{AB} \approx 0$ between QSO-A and QSO-B, whereas the observed \emph{Gaia} value is $\Delta G_{AB} \approx 0.4$. We note that by using the QSOs' flux ratio as an additional model constraint, mass models with high ellipticities are preferred in the optimization. 
In Fig.~\ref{fig::cluster}, we also show a green dotted rectangle representing the approximate region where the third image is predicted by the lens model, based on the positional distribution of 5000 MCMC samples of the posterior parameters.

We also carried out an independent search for the third image candidates by selecting point-like sources with colours consistent with those of the QSO-A and B in the four-dimensional colour space ($g-r$, $r-i$, $i-z$, $z-y$), using the DES DR2 5-band photometric catalogue~\citep{desdr2}. This procedure led us to identify four possible candidates (see magenta squares in Fig.~\ref{fig::cluster}), one of which (labelled ID5) has the smallest r.m.s.~colour difference in the four-dimensional colour space ($\Delta_m  = 0.08 \pm 0.06$) with respect to the QSOs, and appears unresolved with the $1.2''$ seeing conditions. 
If the third image were confirmed through follow-up spectroscopic observations within the predicted region shown in Fig.~\ref{fig::cluster}, QSO~0400--27 may qualify as one of the largest separation strongly-lensed QSO ever observed.

However, two critical aspects emerge from the predictions based on this poorly constrained model (four free parameters and two constraints): i) the third image is predicted to be $1.1\pm 0.3\,\mathrm{mag}$ fainter than QSO-A, whereas ID5, with $g_{\mathrm{DES}} \simeq 22.6$, appears to be $\approx 2.4$ mag fainter ($g_{\mathrm{DES, QSO-A}}\simeq20.2$), making it a less convincing candidate; ii) the encircled cluster mass within $30''$ ($\approx 205$ kpc at $z=0.63$) is predicted from the lens model to be $(3 \pm 1) \, \times 10^{14} \, M_\odot$, which would make it particularly massive when compared with the DES richness parameter~\citep{wen2024}, and more importantly with independent limits which can be inferred from X-ray and SZ surveys. 
We further discuss this point in the next section.


\section{Discussion}
\label{sec::discussion}

Even though the strong similarities between the two QSO spectra support the identification of this system as two multiple images of the same background QSO, lying behind a relatively rich galaxy cluster at $z=0.626$ (see BCG spectrum in ~Fig.~\ref{fig::spectrabcg}), we discuss here the possibility for the QSO~0400--27 system to be a projected physical association of two QSOs. In the simplistic case of fully neglecting the deflection contribution of the intervening lens cluster, the observed two-dimensional $\sim 18''$ separation would correspond to a maximum projected physical distance of $\approx 150$ kpc, at $z=1.345$, for two distinct QSOs. 
Several studies have been devoted to the search for dual AGN over a large range of physical separations, from kiloparsec scale~\citep{chen2022, mannucci2023}, to $>\SI{10}{kpc}$ up to 150 kpc~\citep{hennawi2006, liu2011, eftekharzadeh2017}. A few dual QSOs with velocity differences ($<\SI{100}{km/s}$) and separations ($\sim\! 100$--$150$ kpc) comparable with our system have been identified, however, their spectra show noticeable differences, contrary to what we observe in our case. The system with the most similar characteristics to QSO~0400--27 is SDSS~J1010+0416~\citep{hennawi2006}, consisting of two QSOs at $z=1.52$ separated by $\approx \SI{150}{kpc}$. The rest frame velocity difference of SDSS~J1010+0416 based on the MgII emission line is $\approx \SI{40}{km/s}$, even though the two MgII lines appear to have different widths, while the CIII lines show a velocity offset of $\approx \SI{2000}{km/s}$. No shifts are detected for the CIII and MgII lines in our case ($\Delta v_{\rm rest}\lesssim 40\, {\rm km/s}$). 

As mentioned above, a noticeable tension seems to exist when comparing the encircled lensing mass to $M_{500}$ estimated from optical mass proxies. Such comparisons, however, are affected by large systematic errors due to the uncertain extrapolation of the SL mass to much larger radii and large uncertainties in the cluster mass calibration from richness parameters. To this respect, a more meaningful test is to check whether X-ray emission is expected in the eROSITA~\citep{predehl2021} survey data, based on the estimated lensing mass of WHJ0400--27. To this end, we inspected the eRASS catalogue of 12\,247 clusters published in~\citet{bulbul2024}. Our system is not detected, nor the two QSOs, which would be blended in the eROSITA PSF. By estimating a lower limit for the X-ray luminosity ($L_X\,\rm{[0.2-2.3\, keV ]} \approx 10^{44}\, {\rm erg\,s^{-1}}$, within their adopted $R_{500}$) in the vicinity of WHJ0400--27, using the flux limit in that survey region and the cluster redshift, we estimate that a cluster with a mass in excess of $M_{500} \approx 5 \times 10^{14}\, M_\odot$ should be detected in the eRASS data. This seems to be in tension with the extrapolated SL mass to $R_{500}$, which we estimate to be $M_{500}\approx 8\times 10^{14}\, M_\odot$ using the best-fit mass-density profile from our lens model. However, these estimates are both affected by large systematic uncertainties.

As an additional interesting argument, we note that no detection of our cluster is reported in the ACT Sunyaev-Zeldovich (SZ) survey from the DR5 MCMF cluster catalogue~\citep{klein2024}, where several clusters at similar redshifts have been identified with masses down to $M_{500} \approx 2 \times 10^{14} \, M_\odot$. Both the X-ray and SZ-based mass lower limits seem to indicate a tension with the lens model based mass, which at face value is a factor $\approx 2\text{--}4$, notwithstanding all the uncertainties in the quoted mass values. 

\section{Conclusions}
\label{sec::conclusions}

As part of a search of gravitationally lensed \emph{Gaia} QSOs by galaxy clusters, we present the discovery of a pair of QSOs separated by $\Delta\vartheta=17.8''$ at $z_{\rm QSO, spec}=1.345$ with remarkable spectral similarities, lying in the projected vicinity of the galaxy cluster candidate WHJ0400--27 at $z_{\rm CL, phot}=0.63$. Our two independent spectroscopic studies, carried out at the ESO/NTT and NOT telescopes, show indistinguishable spectra for the two QSOs, with no measurable redshift difference, below the third decimal digit, 
very similar continuum, emission line shapes and line flux ratios (cfr.~Fig.~\ref{fig::spectra}).

We first investigated the SL scenario by building a model of the optically-selected cluster, using a single-component mass distribution centred on the BCG ($z_{\rm BCG, spec}=0.626$). We used the position of the two QSOs, at a projected distance of $31.5''$ and $24.5''$ from the BCG, as the only model constraint.  Such model predicts the existence of a third lensed image, which remains to be found among a number of candidates selected from sources with similar morphology and colours. Notably, our poorly constrained lensing model predicts a cluster mass that appears to be a factor 2 to 4 larger than mass upper limits obtained from the lack of X-ray and SZ detections from eROSITA and ACT surveys, respectively. Even though systematic uncertainties remain large when comparing the SL mass of WHJ0400--27 with that inferred from other mass proxies (richness, X-rays, SZ), this raises questions about the lensing nature of the QSO pair. Interestingly, several other strong lensing systems on galaxy scale face comparable challenges (e.g.~\citealt{anguita2018, lemon2022}). 
It is therefore worth exploring the possibility that QSO~0400--27 represents a physical association of two QSOs, separated by approximately $150\,\mathrm{kpc}$ in projection, behind WHJ0400--27 but outside the SL regime. While dual AGN with such large separations and small velocity shifts are rare, the few observed have been found with noticeable differences in their spectra (e.g.~line widths, see~\citealt{hennawi2006}). We therefore conclude that the combination of physical separation and nearly identical spectra would make our dual QSO system rather unique to date. 




Planned follow-up spectroscopic and imaging observations  will be crucial to significantly improve the modelling constraints of the lens cluster, specifically by identifying the predicted third image and by confirming a sizable number of cluster galaxies. Should we be able to confirm the gravitational lensing nature of our system, QSO~0400--27 has the potential to become the largest-separation lensed QSO known to date, holding great promise for TDC applications. 


\begin{acknowledgements}
We thank the anonymous referee for the helpful and insightful comments that improved the paper.\\
LB is indebted to the communities behind the multiple free, libre, and open-source software packages on which we all depend on.\\
PR conceived the research; LB, GA, and PR developed the methodology; LB, GA, MS, PB and GDR performed all the numerical work, including software development, investigation, and validation. All authors contributed to the discussion of the results, the editing and revision of the paper.\\
We acknowledge support from the Italian Ministry of University and Research through grant PRIN-MIUR 2020SKSTHZ.\\
This publication  was produced while attending the PhD program in PhD in Space Science and Technology at the University of Trento, Cycle XXXIX, with the support of a scholarship financed by the Ministerial Decree no.~118 of 2nd march 2023, based on the NRRP - funded by the European Union - NextGenerationEU - Mission 4 ``Education and Research'', Component 1 ``Enhancement of the offer of educational services: from nurseries to universities'' - Investment 4.1 ``Extension of the number of research doctorates and innovative doctorates for public administration and cultural heritage''.\\
Based on observations made with the Nordic Optical Telescope, owned in collaboration by the University of Turku and Aarhus University, and operated jointly by Aarhus University, the University of Turku and the University of Oslo, representing Denmark, Finland and Norway, the University of Iceland and Stockholm University at the Observatorio del Roque de los Muchachos, La Palma, Spain, of the Instituto de Astrofisica de Canarias.\\
The data presented here were obtained [in part] with ALFOSC, which is provided by the Instituto de Astrofisica de Andalucia (IAA) under a joint agreement with the University of Copenhagen and NOT.\\
Based on observations collected at the European Organisation for Astronomical Research in the Southern Hemisphere under ESO programme 112.25CT.\\

This work uses the following software packages:
    \href{https://www.python.org/}{\texttt{Python}}
    \citep{python},
    \href{https://github.com/numpy/numpy}{\texttt{NumPy}}
    \citep{numpy1, numpy2},
    \href{https://github.com/scipy/scipy}{\texttt{SciPy}}
    \citep{scipy},
    \href{https://github.com/astropy/astropy}{\texttt{Astropy}}
    \citep{astropy1, astropy2},
    \href{https://github.com/matplotlib/matplotlib}{\texttt{matplotlib}}
    \citep{matplotlib},
    \href{https://github.com/pypeit/PypeIt}{\texttt{PypeIt}}
    \citep{prochaska20},
    \href{}{\texttt{Gravity.jl}}
    \citep{lombardi2024},
    \href{https://git-cral.univ-lyon1.fr/lenstool/lenstool}{\texttt{lenstool}}
    \citep{kneib1996, jullo2007, jullo2009},
    \href{https://automeris.io/WebPlotDigitizer.html}{\texttt{WebPlotDigitizer}}
    \citep{rohatgi2024},
    \href{https://sites.google.com/cfa.harvard.edu/saoimageds9/home}{\texttt{ds9}}
    \citep{ds9},
    \href{https://aladin.cds.unistra.fr/}{\texttt{Aladin}}
    \citep{aladin},
    \href{https://www.star.bris.ac.uk/~mbt/topcat/}{\texttt{topcat}}
    \citep{topcat},
    \href{https://www.gnu.org/software/bash/}{\texttt{bash}}
    \citep{gnu2007free}.
\end{acknowledgements}

%
%

\bibliographystyle{aa}
\bibliography{bibliography}

\begin{thebibliography}{61}
\expandafter\ifx\csname natexlab\endcsname\relax\def\natexlab#1{#1}\fi

\bibitem[{{Abbott} {et~al.}(2021){Abbott}, {Adam{\'o}w}, {Aguena}, {Allam},
  {Amon}, {Annis}, {Avila}, {Bacon}, {Banerji}, {Bechtol}, {Becker},
  {Bernstein}, {Bertin}, {Bhargava}, {Bridle}, {Brooks}, {Burke}, {Carnero
  Rosell}, {Carrasco Kind}, {Carretero}, {Castander}, {Cawthon}, {Chang},
  {Choi}, {Conselice}, {Costanzi}, {Crocce}, {da Costa}, {Davis}, {De Vicente},
  {DeRose}, {Desai}, {Diehl}, {Dietrich}, {Drlica-Wagner}, {Eckert},
  {Elvin-Poole}, {Everett}, {Evrard}, {Ferrero}, {Fert{\'e}}, {Flaugher},
  {Fosalba}, {Friedel}, {Frieman}, {Garc{\'\i}a-Bellido}, {Gaztanaga},
  {Gelman}, {Gerdes}, {Giannantonio}, {Gill}, {Gruen}, {Gruendl}, {Gschwend},
  {Gutierrez}, {Hartley}, {Hinton}, {Hollowood}, {Honscheid}, {Huterer},
  {James}, {Jeltema}, {Johnson}, {Kent}, {Kron}, {Kuehn}, {Kuropatkin},
  {Lahav}, {Li}, {Lidman}, {Lin}, {MacCrann}, {Maia}, {Manning}, {Maloney},
  {March}, {Marshall}, {Martini}, {Melchior}, {Menanteau}, {Miquel}, {Morgan},
  {Myles}, {Neilsen}, {Ogando}, {Palmese}, {Paz-Chinch{\'o}n}, {Petravick},
  {Pieres}, {Plazas}, {Pond}, {Rodriguez-Monroy}, {Romer}, {Roodman}, {Rykoff},
  {Sako}, {Sanchez}, {Santiago}, {Scarpine}, {Serrano}, {Sevilla-Noarbe},
  {Smith}, {Smith}, {Soares-Santos}, {Suchyta}, {Swanson}, {Tarle}, {Thomas},
  {To}, {Tremblay}, {Troxel}, {Tucker}, {Turner}, {Varga}, {Walker},
  {Wechsler}, {Weller}, {Wester}, {Wilkinson}, {Yanny}, {Zhang}, {Nikutta},
  {Fitzpatrick}, {Jacques}, {Scott}, {Olsen}, {Huang}, {Herrera}, {Juneau},
  {Nidever}, {Weaver}, {Adean}, {Correia}, {de Freitas}, {Freitas},
  {Singulani}, {Vila-Verde}, \& {Linea Science Server}}]{desdr2}
{Abbott}, T.~M.~C., {Adam{\'o}w}, M., {Aguena}, M., {et~al.} 2021, \apjs, 255,
  20

\bibitem[{{Anguita} {et~al.}(2018){Anguita}, {Schechter}, {Kuropatkin},
  {Morgan}, {Ostrovski}, {Abramson}, {Agnello}, {Apostolovski}, {Fassnacht},
  {Hsueh}, {Motta}, {Rojas}, {Rusu}, {Treu}, {Williams}, {Auger},
  {Buckley-Geer}, {Lin}, {McMahon}, {Abbott}, {Allam}, {Annis}, {Bernstein},
  {Bertin}, {Brooks}, {Burke}, {Carnero Rosell}, {Carrasco-Kind}, {Carretero},
  {Cunha}, {D'Andrea}, {De Vicente}, {DePoy}, {Desai}, {Diehl}, {Doel},
  {Flaugher}, {Garc{\'\i}a-Bellido}, {Gerdes}, {Gruen}, {Gruendl}, {Gschwend},
  {Hartley}, {Hollowood}, {Honscheid}, {James}, {Kuehn}, {Lima}, {Maia},
  {Miquel}, {Plazas}, {Sanchez}, {Scarpine}, {Smith}, {Soares-Santos},
  {Sobreira}, {Suchyta}, {Tarle}, \& {Walker}}]{anguita2018}
{Anguita}, T., {Schechter}, P.~L., {Kuropatkin}, N., {et~al.} 2018, \mnras,
  480, 5017

\bibitem[{{Astropy Collaboration} {et~al.}(2013){Astropy Collaboration},
  {Robitaille}, {Tollerud}, {Greenfield}, {Droettboom}, {Bray}, {Aldcroft},
  {Davis}, {Ginsburg}, {Price-Whelan}, {Kerzendorf}, {Conley}, {Crighton},
  {Barbary}, {Muna}, {Ferguson}, {Grollier}, {Parikh}, {Nair}, {Unther},
  {Deil}, {Woillez}, {Conseil}, {Kramer}, {Turner}, {Singer}, {Fox}, {Weaver},
  {Zabalza}, {Edwards}, {Azalee Bostroem}, {Burke}, {Casey}, {Crawford},
  {Dencheva}, {Ely}, {Jenness}, {Labrie}, {Lim}, {Pierfederici}, {Pontzen},
  {Ptak}, {Refsdal}, {Servillat}, \& {Streicher}}]{astropy1}
{Astropy Collaboration}, {Robitaille}, T.~P., {Tollerud}, E.~J., {et~al.} 2013,
  \aap, 558, A33

\bibitem[{{Birrer} {et~al.}(2016){Birrer}, {Amara}, \&
  {Refregier}}]{birrer2016}
{Birrer}, S., {Amara}, A., \& {Refregier}, A. 2016, \jcap, 2016, 020

\bibitem[{{Bonnarel} {et~al.}(2000){Bonnarel}, {Fernique}, {Bienaym{\'e}},
  {Egret}, {Genova}, {Louys}, {Ochsenbein}, {Wenger}, \& {Bartlett}}]{aladin}
{Bonnarel}, F., {Fernique}, P., {Bienaym{\'e}}, O., {et~al.} 2000, \aaps, 143,
  33

\bibitem[{{Bulbul} {et~al.}(2024){Bulbul}, {Liu}, {Kluge}, {Zhang}, {Sanders},
  {Bahar}, {Ghirardini}, {Artis}, {Seppi}, {Garrel}, {Ramos-Ceja}, {Comparat},
  {Balzer}, {B{\"o}ckmann}, {Br{\"u}ggen}, {Clerc}, {Dennerl}, {Dolag},
  {Freyberg}, {Grandis}, {Gruen}, {Kleinebreil}, {Krippendorf}, {Lamer},
  {Merloni}, {Migkas}, {Nandra}, {Pacaud}, {Predehl}, {Reiprich}, {Schrabback},
  {Veronica}, {Weller}, \& {Zelmer}}]{bulbul2024}
{Bulbul}, E., {Liu}, A., {Kluge}, M., {et~al.} 2024, \aap, 685, A106

\bibitem[{{Buzzoni} {et~al.}(1984){Buzzoni}, {Delabre}, {Dekker}, {Dodorico},
  {Enard}, {Focardi}, {Gustafsson}, {Nees}, {Paureau}, \&
  {Reiss}}]{buzzoni1984}
{Buzzoni}, B., {Delabre}, B., {Dekker}, H., {et~al.} 1984, The Messenger, 38, 9

\bibitem[{{Chen} {et~al.}(2022){Chen}, {Hwang}, {Shen}, {Liu}, {Zakamska},
  {Yang}, \& {Li}}]{chen2022}
{Chen}, Y.-C., {Hwang}, H.-C., {Shen}, Y., {et~al.} 2022, \apj, 925, 162

\bibitem[{{Dahle} {et~al.}(2013){Dahle}, {Gladders}, {Sharon}, {Bayliss},
  {Wuyts}, {Abramson}, {Koester}, {Groeneboom}, {Brinckmann}, {Kristensen},
  {Lindholmer}, {Nielsen}, {Krogager}, \& {Fynbo}}]{dahle2013}
{Dahle}, H., {Gladders}, M.~D., {Sharon}, K., {et~al.} 2013, \apj, 773, 146

\bibitem[{{Dark Energy Survey Collaboration} {et~al.}(2016){Dark Energy Survey
  Collaboration}, {Abbott}, {Abdalla}, {Aleksi{\'c}}, {Allam}, {Amara},
  {Bacon}, {Balbinot}, {Banerji}, {Bechtol}, {Benoit-L{\'e}vy}, {Bernstein},
  {Bertin}, {Blazek}, {Bonnett}, {Bridle}, {Brooks}, {Brunner}, {Buckley-Geer},
  {Burke}, {Caminha}, {Capozzi}, {Carlsen}, {Carnero-Rosell}, {Carollo},
  {Carrasco-Kind}, {Carretero}, {Castander}, {Clerkin}, {Collett}, {Conselice},
  {Crocce}, {Cunha}, {D'Andrea}, {da Costa}, {Davis}, {Desai}, {Diehl},
  {Dietrich}, {Dodelson}, {Doel}, {Drlica-Wagner}, {Estrada}, {Etherington},
  {Evrard}, {Fabbri}, {Finley}, {Flaugher}, {Foley}, {Fosalba}, {Frieman},
  {Garc{\'\i}a-Bellido}, {Gaztanaga}, {Gerdes}, {Giannantonio}, {Goldstein},
  {Gruen}, {Gruendl}, {Guarnieri}, {Gutierrez}, {Hartley}, {Honscheid}, {Jain},
  {James}, {Jeltema}, {Jouvel}, {Kessler}, {King}, {Kirk}, {Kron}, {Kuehn},
  {Kuropatkin}, {Lahav}, {Li}, {Lima}, {Lin}, {Maia}, {Makler}, {Manera},
  {Maraston}, {Marshall}, {Martini}, {McMahon}, {Melchior}, {Merson}, {Miller},
  {Miquel}, {Mohr}, {Morice-Atkinson}, {Naidoo}, {Neilsen}, {Nichol}, {Nord},
  {Ogando}, {Ostrovski}, {Palmese}, {Papadopoulos}, {Peiris}, {Peoples},
  {Percival}, {Plazas}, {Reed}, {Refregier}, {Romer}, {Roodman}, {Ross},
  {Rozo}, {Rykoff}, {Sadeh}, {Sako}, {S{\'a}nchez}, {Sanchez}, {Santiago},
  {Scarpine}, {Schubnell}, {Sevilla-Noarbe}, {Sheldon}, {Smith}, {Smith},
  {Soares-Santos}, {Sobreira}, {Soumagnac}, {Suchyta}, {Sullivan}, {Swanson},
  {Tarle}, {Thaler}, {Thomas}, {Thomas}, {Tucker}, {Vieira}, {Vikram},
  {Walker}, {Wechsler}, {Weller}, {Wester}, {Whiteway}, {Wilcox}, {Yanny},
  {Zhang}, \& {Zuntz}}]{des2016}
{Dark Energy Survey Collaboration}, {Abbott}, T., {Abdalla}, F.~B., {et~al.}
  2016, \mnras, 460, 1270

\bibitem[{{Djupvik} \& {Andersen}(2010)}]{djupvik2010}
{Djupvik}, A.~A. \& {Andersen}, J. 2010, in Astrophysics and Space Science
  Proceedings, Vol.~14, Highlights of Spanish Astrophysics V, 211

\bibitem[{{Dutta} {et~al.}(2024){Dutta}, {Acebron}, {Fumagalli}, {Grillo},
  {Caminha}, \& {Fossati}}]{dutta2024}
{Dutta}, R., {Acebron}, A., {Fumagalli}, M., {et~al.} 2024, \mnras, 528, 1895

\bibitem[{{Eftekharzadeh} {et~al.}(2017){Eftekharzadeh}, {Myers}, {Hennawi},
  {Djorgovski}, {Richards}, {Mahabal}, \& {Graham}}]{eftekharzadeh2017}
{Eftekharzadeh}, S., {Myers}, A.~D., {Hennawi}, J.~F., {et~al.} 2017, \mnras,
  468, 77

\bibitem[{{Euclid Collaboration} {et~al.}(2019){Euclid Collaboration}, {Adam},
  {Vannier}, {Maurogordato}, {Biviano}, {Adami}, {Ascaso}, {Bellagamba},
  {Benoist}, {Cappi}, {D{\'\i}az-S{\'a}nchez}, {Durret}, {Farrens}, {Gonzalez},
  {Iovino}, {Licitra}, {Maturi}, {Mei}, {Merson}, {Munari}, {Pell{\'o}},
  {Ricci}, {Rocci}, {Roncarelli}, {Sarron}, {Amoura}, {Andreon}, {Apostolakos},
  {Arnaud}, {Bardelli}, {Bartlett}, {Baugh}, {Borgani}, {Brodwin}, {Castander},
  {Castignani}, {Cucciati}, {De Lucia}, {Dubath}, {Fosalba}, {Giocoli},
  {Hoekstra}, {Mamon}, {Melin}, {Moscardini}, {Paltani}, {Radovich},
  {Sartoris}, {Schultheis}, {Sereno}, {Weller}, {Burigana}, {Carvalho},
  {Corcione}, {Kurki-Suonio}, {Lilje}, {Sirri}, {Toledo-Moreo}, \&
  {Zamorani}}]{euclid2019}
{Euclid Collaboration}, {Adam}, R., {Vannier}, M., {et~al.} 2019, \aap, 627,
  A23

\bibitem[{{Filippenko}(1982)}]{filippenko1982}
{Filippenko}, A.~V. 1982, \pasp, 94, 715

\bibitem[{{Gaia Collaboration} {et~al.}(2023){Gaia Collaboration},
  {Bailer-Jones}, {Teyssier}, {Delchambre}, {Ducourant}, {Garabato},
  {Hatzidimitriou}, {Klioner}, {Rimoldini}, {Bellas-Velidis}, {Carballo},
  {Carnerero}, {Diener}, {Fouesneau}, {Galluccio}, {Gavras}, {Krone-Martins},
  {Raiteri}, {Teixeira}, {Brown}, {Vallenari}, {Prusti}, {de Bruijne},
  {Arenou}, {Babusiaux}, {Biermann}, {Creevey}, {Evans}, {Eyer}, {Guerra},
  {Hutton}, {Jordi}, {Lammers}, {Lindegren}, {Luri}, {Mignard}, {Panem},
  {Pourbaix}, {Randich}, {Sartoretti}, {Soubiran}, {Tanga}, {Walton},
  {Bastian}, {Drimmel}, {Jansen}, {Katz}, {Lattanzi}, {van Leeuwen}, {Bakker},
  {Cacciari}, {Casta{\~n}eda}, {De Angeli}, {Fabricius}, {Fr{\'e}mat},
  {Guerrier}, {Heiter}, {Masana}, {Messineo}, {Mowlavi}, {Nicolas},
  {Nienartowicz}, {Pailler}, {Panuzzo}, {Riclet}, {Roux}, {Seabroke}, {Sordo},
  {Th{\'e}venin}, {Gracia-Abril}, {Portell}, {Altmann}, {Andrae}, {Audard},
  {Benson}, {Berthier}, {Blomme}, {Burgess}, {Busonero}, {Busso},
  {C{\'a}novas}, {Carry}, {Cellino}, {Cheek}, {Clementini}, {Damerdji},
  {Davidson}, {de Teodoro}, {Nu{\~n}ez Campos}, {Dell'Oro}, {Esquej},
  {Fern{\'a}ndez-Hern{\'a}ndez}, {Fraile}, {Garc{\'\i}a-Lario}, {Gosset},
  {Haigron}, {Halbwachs}, {Hambly}, {Harrison}, {Hern{\'a}ndez}, {Hestroffer},
  {Hodgkin}, {Holl}, {Jan{\ss}en}, {Jevardat de Fombelle}, {Jordan},
  {Lanzafame}, {L{\"o}ffler}, {Marchal}, {Marrese}, {Moitinho}, {Muinonen},
  {Osborne}, {Pancino}, {Pauwels}, {Recio-Blanco}, {Reyl{\'e}}, {Riello},
  {Roegiers}, {Rybizki}, {Sarro}, {Siopis}, {Smith}, {Sozzetti}, {Utrilla},
  {van Leeuwen}, {Abbas}, {{\'A}brah{\'a}m}, {Abreu Aramburu}, {Aerts},
  {Aguado}, {Ajaj}, {Aldea-Montero}, {Altavilla}, {{\'A}lvarez}, {Alves},
  {Anderson}, {Anglada Varela}, {Antoja}, {Baines}, {Baker},
  {Balaguer-N{\'u}{\~n}ez}, {Balbinot}, {Balog}, {Barache}, {Barbato},
  {Barros}, {Barstow}, {Bartolom{\'e}}, {Bassilana}, {Bauchet}, {Becciani},
  {Bellazzini}, {Berihuete}, {Bernet}, {Bertone}, {Bianchi}, {Binnenfeld},
  {Blanco-Cuaresma}, {Boch}, {Bombrun}, {Bossini}, {Bouquillon}, {Bragaglia},
  {Bramante}, {Breedt}, {Bressan}, {Brouillet}, {Brugaletta}, {Bucciarelli},
  {Burlacu}, {Butkevich}, {Buzzi}, {Caffau}, {Cancelliere}, {Cantat-Gaudin},
  {Carlucci}, {Carrasco}, {Casamiquela}, {Castellani}, {Castro-Ginard},
  {Chaoul}, {Charlot}, {Chemin}, {Chiaramida}, {Chiavassa}, {Chornay},
  {Comoretto}, {Contursi}, {Cooper}, {Cornez}, {Cowell}, {Crifo}, {Cropper},
  {Crosta}, {Crowley}, {Dafonte}, {Dapergolas}, {David}, {de Laverny}, {De
  Luise}, {De March}, {De Ridder}, {de Souza}, {de Torres}, {del Peloso}, {del
  Pozo}, {Delbo}, {Delgado}, {Delisle}, {Demouchy}, {Dharmawardena}, {Diakite},
  {Distefano}, {Dolding}, {Enke}, {Fabre}, {Fabrizio}, {Faigler}, {Fedorets},
  {Fernique}, {Figueras}, {Fournier}, {Fouron}, {Fragkoudi}, {Gai},
  {Garcia-Gutierrez}, {Garcia-Reinaldos}, {Garc{\'\i}a-Torres}, {Garofalo},
  {Gavel}, {Gerlach}, {Geyer}, {Giacobbe}, {Gilmore}, {Girona}, {Giuffrida},
  {Gomel}, {Gomez}, {Gonz{\'a}lez-N{\'u}{\~n}ez},
  {Gonz{\'a}lez-Santamar{\'\i}a}, {Gonz{\'a}lez-Vidal}, {Granvik}, {Guillout},
  {Guiraud}, {Guti{\'e}rrez-S{\'a}nchez}, {Guy}, {Hauser}, {Haywood}, {Helmer},
  {Helmi}, {Sarmiento}, {Hidalgo}, {Hilger}, {H{\l}adczuk}, {Hobbs}, {Holland},
  {Huckle}, {Jardine}, {Jasniewicz}, {Jean-Antoine Piccolo},
  {Jim{\'e}nez-Arranz}, {Juaristi Campillo}, {Julbe}, {Karbevska}, {Kervella},
  {Khanna}, {Kontizas}, {Kordopatis}, {Korn}, {K{\'o}sp{\'a}l},
  {Kostrzewa-Rutkowska}, {Kruszy{\'n}ska}, {Kun}, {Laizeau}, {Lambert},
  {Lanza}, {Lasne}, {Le Campion}, {Lebreton}, {Lebzelter}, {Leccia}, {Leclerc},
  {Lecoeur-Taibi}, {Liao}, {Licata}, {Lindstr{\o}m}, {Lister}, {Livanou},
  {Lobel}, {Lorca}, {Loup}, {Madrero Pardo}, {Magdaleno Romeo}, {Managau},
  {Mann}, {Manteiga}, {Marchant}, {Marconi}, {Marcos}, {Marcos Santos},
  {Mar{\'\i}n Pina}, {Marinoni}, {Marocco}, {Marshall}, {Martin Polo},
  {Mart{\'\i}n-Fleitas}, {Marton}, {Mary}, {Masip}, {Massari},
  {Mastrobuono-Battisti}, {Mazeh}, {McMillan}, {Messina}, {Michalik}, {Millar},
  {Mints}, {Molina}, {Molinaro}, {Moln{\'a}r}, {Monari}, {Mongui{\'o}},
  {Montegriffo}, {Montero}, {Mor}, {Mora}, {Morbidelli}, {Morel}, {Morris},
  {Muraveva}, {Murphy}, {Musella}, {Nagy}, {Noval}, {Oca{\~n}a}, {Ogden},
  {Ordenovic}, {Osinde}, {Pagani}, {Pagano}, {Palaversa}, {Palicio},
  {Pallas-Quintela}, {Panahi}, {Payne-Wardenaar}, {Pe{\~n}alosa Esteller},
  {Penttil{\"a}}, {Pichon}, {Piersimoni}, {Pineau}, {Plachy}, {Plum}, {Poggio},
  {Pr{\v{s}}a}, {Pulone}, {Racero}, {Ragaini}, {Rainer}, {Ramos},
  {Ramos-Lerate}, {Re Fiorentin}, {Regibo}, {Richards}, {Rios Diaz}, {Ripepi},
  {Riva}, {Rix}, {Rixon}, {Robichon}, {Robin}, {Robin}, {Roelens}, {Rogues},
  {Rohrbasser}, {Romero-G{\'o}mez}, {Rowell}, {Royer}, {Ruz Mieres}, {Rybicki},
  {Sadowski}, {S{\'a}ez N{\'u}{\~n}ez}, {Sagrist{\`a} Sell{\'e}s}, {Sahlmann},
  {Salguero}, {Samaras}, {Sanchez Gimenez}, {Sanna}, {Santove{\~n}a},
  {Sarasso}, {Schultheis}, {Sciacca}, {Segol}, {Segovia}, {S{\'e}gransan},
  {Semeux}, {Shahaf}, {Siddiqui}, {Siebert}, {Siltala}, {Silvelo}, {Slezak},
  {Slezak}, {Smart}, {Snaith}, {Solano}, {Solitro}, {Souami}, {Souchay},
  {Spagna}, {Spina}, {Spoto}, {Steele}, {Steidelm{\"u}ller}, {Stephenson},
  {S{\"u}veges}, {Surdej}, {Szabados}, {Szegedi-Elek}, {Taris}, {Taylor},
  {Tolomei}, {Tonello}, {Torra}, {Torra}, {Torralba Elipe}, {Trabucchi},
  {Tsounis}, {Turon}, {Ulla}, {Unger}, {Vaillant}, {van Dillen}, {van Reeven},
  {Vanel}, {Vecchiato}, {Viala}, {Vicente}, {Voutsinas}, {Weiler}, {Wevers},
  {Wyrzykowski}, {Yoldas}, {Yvard}, {Zhao}, {Zorec}, {Zucker}, \&
  {Zwitter}}]{gaiadr3qso}
{Gaia Collaboration}, {Bailer-Jones}, C.~A.~L., {Teyssier}, D., {et~al.} 2023,
  \aap, 674, A41

\bibitem[{{Garilli} {et~al.}(2010){Garilli}, {Fumana}, {Franzetti}, {Paioro},
  {Scodeggio}, {Le F{\`e}vre}, {Paltani}, \& {Scaramella}}]{garilli2010}
{Garilli}, B., {Fumana}, M., {Franzetti}, P., {et~al.} 2010, \pasp, 122, 827

\bibitem[{GNU(2007)}]{gnu2007free}
GNU, P. 2007, Free Software Foundation. Bash (3.2. 48)[Unix shell program]

\bibitem[{{Grillo} {et~al.}(2024){Grillo}, {Pagano}, {Rosati}, \&
  {Suyu}}]{grillo2024}
{Grillo}, C., {Pagano}, L., {Rosati}, P., \& {Suyu}, S.~H. 2024, \aap, 684, L23

\bibitem[{{Grillo} {et~al.}(2018){Grillo}, {Rosati}, {Suyu}, {Balestra},
  {Caminha}, {Halkola}, {Kelly}, {Lombardi}, {Mercurio}, {Rodney}, \&
  {Treu}}]{grillo2018}
{Grillo}, C., {Rosati}, P., {Suyu}, S.~H., {et~al.} 2018, \apj, 860, 94

\bibitem[{{Grillo} {et~al.}(2020){Grillo}, {Rosati}, {Suyu}, {Caminha},
  {Mercurio}, \& {Halkola}}]{grillo2020}
{Grillo}, C., {Rosati}, P., {Suyu}, S.~H., {et~al.} 2020, \apj, 898, 87

\bibitem[{Harris {et~al.}(2020)Harris, Millman, van~der Walt, Gommers,
  Virtanen, Cournapeau, Wieser, Taylor, Berg, Smith, Kern, Picus, Hoyer, van
  Kerkwijk, Brett, Haldane, Fernández~del Río, Wiebe, Peterson,
  Gérard-Marchant, Sheppard, Reddy, Weckesser, Abbasi, Gohlke, \&
  Oliphant}]{numpy2}
Harris, C.~R., Millman, K.~J., van~der Walt, S.~J., {et~al.} 2020, Nature, 585,
  357

\bibitem[{{Hennawi} {et~al.}(2006){Hennawi}, {Strauss}, {Oguri}, {Inada},
  {Richards}, {Pindor}, {Schneider}, {Becker}, {Gregg}, {Hall}, {Johnston},
  {Fan}, {Burles}, {Schlegel}, {Gunn}, {Lupton}, {Bahcall}, {Brunner}, \&
  {Brinkmann}}]{hennawi2006}
{Hennawi}, J.~F., {Strauss}, M.~A., {Oguri}, M., {et~al.} 2006, \aj, 131, 1

\bibitem[{Hunter(2007)}]{matplotlib}
Hunter, J.~D. 2007, Computing in Science \& Engineering, 9, 90

\bibitem[{{Inada} {et~al.}(2006){Inada}, {Oguri}, {Morokuma}, {Doi}, {Yasuda},
  {Becker}, {Richards}, {Kochanek}, {Kayo}, {Konishi}, {Utsunomiya}, {Shin},
  {Strauss}, {Sheldon}, {York}, {Hennawi}, {Schneider}, {Dai}, \&
  {Fukugita}}]{inada2006}
{Inada}, N., {Oguri}, M., {Morokuma}, T., {et~al.} 2006, \apjl, 653, L97

\bibitem[{{Inada} {et~al.}(2003){Inada}, {Oguri}, {Pindor}, {Hennawi}, {Chiu},
  {Zheng}, {Ichikawa}, {Gregg}, {Becker}, {Suto}, {Strauss}, {Turner},
  {Keeton}, {Annis}, {Castander}, {Eisenstein}, {Frieman}, {Fukugita}, {Gunn},
  {Johnston}, {Kent}, {Nichol}, {Richards}, {Rix}, {Sheldon}, {Bahcall},
  {Brinkmann}, {Ivezi{\'c}}, {Lamb}, {McKay}, {Schneider}, \&
  {York}}]{inada2003}
{Inada}, N., {Oguri}, M., {Pindor}, B., {et~al.} 2003, \nat, 426, 810

\bibitem[{{Ivezi{\'c}} {et~al.}(2019){Ivezi{\'c}}, {Kahn}, {Tyson}, {Abel},
  {Acosta}, {Allsman}, {Alonso}, {AlSayyad}, {Anderson}, {Andrew}, {Angel},
  {Angeli}, {Ansari}, {Antilogus}, {Araujo}, {Armstrong}, {Arndt}, {Astier},
  {Aubourg}, {Auza}, {Axelrod}, {Bard}, {Barr}, {Barrau}, {Bartlett}, {Bauer},
  {Bauman}, {Baumont}, {Bechtol}, {Bechtol}, {Becker}, {Becla}, {Beldica},
  {Bellavia}, {Bianco}, {Biswas}, {Blanc}, {Blazek}, {Bland ford}, {Bloom},
  {Bogart}, {Bond}, {Booth}, {Borgland}, {Borne}, {Bosch}, {Boutigny},
  {Brackett}, {Bradshaw}, {Brand t}, {Brown}, {Bullock}, {Burchat}, {Burke},
  {Cagnoli}, {Calabrese}, {Callahan}, {Callen}, {Carlin}, {Carlson}, {Chand
  rasekharan}, {Charles-Emerson}, {Chesley}, {Cheu}, {Chiang}, {Chiang},
  {Chirino}, {Chow}, {Ciardi}, {Claver}, {Cohen-Tanugi}, {Cockrum}, {Coles},
  {Connolly}, {Cook}, {Cooray}, {Covey}, {Cribbs}, {Cui}, {Cutri}, {Daly},
  {Daniel}, {Daruich}, {Daubard}, {Daues}, {Dawson}, {Delgado}, {Dellapenna},
  {de Peyster}, {de Val-Borro}, {Digel}, {Doherty}, {Dubois},
  {Dubois-Felsmann}, {Durech}, {Economou}, {Eifler}, {Eracleous}, {Emmons},
  {Fausti Neto}, {Ferguson}, {Figueroa}, {Fisher-Levine}, {Focke}, {Foss},
  {Frank}, {Freemon}, {Gangler}, {Gawiser}, {Geary}, {Gee}, {Geha}, {Gessner},
  {Gibson}, {Gilmore}, {Glanzman}, {Glick}, {Goldina}, {Goldstein}, {Goodenow},
  {Graham}, {Gressler}, {Gris}, {Guy}, {Guyonnet}, {Haller}, {Harris},
  {Hascall}, {Haupt}, {Hernand ez}, {Herrmann}, {Hileman}, {Hoblitt},
  {Hodgson}, {Hogan}, {Howard}, {Huang}, {Huffer}, {Ingraham}, {Innes},
  {Jacoby}, {Jain}, {Jammes}, {Jee}, {Jenness}, {Jernigan}, {Jevremovi{\'c}},
  {Johns}, {Johnson}, {Johnson}, {Jones}, {Juramy-Gilles}, {Juri{\'c}},
  {Kalirai}, {Kallivayalil}, {Kalmbach}, {Kantor}, {Karst}, {Kasliwal},
  {Kelly}, {Kessler}, {Kinnison}, {Kirkby}, {Knox}, {Kotov}, {Krabbendam},
  {Krughoff}, {Kub{\'a}nek}, {Kuczewski}, {Kulkarni}, {Ku}, {Kurita}, {Lage},
  {Lambert}, {Lange}, {Langton}, {Le Guillou}, {Levine}, {Liang}, {Lim},
  {Lintott}, {Long}, {Lopez}, {Lotz}, {Lupton}, {Lust}, {MacArthur}, {Mahabal},
  {Mand elbaum}, {Markiewicz}, {Marsh}, {Marshall}, {Marshall}, {May},
  {McKercher}, {McQueen}, {Meyers}, {Migliore}, {Miller}, {Mills}, {Miraval},
  {Moeyens}, {Moolekamp}, {Monet}, {Moniez}, {Monkewitz}, {Montgomery},
  {Morrison}, {Mueller}, {Muller}, {Mu{\~n}oz Arancibia}, {Neill}, {Newbry},
  {Nief}, {Nomerotski}, {Nordby}, {O'Connor}, {Oliver}, {Olivier}, {Olsen},
  {O'Mullane}, {Ortiz}, {Osier}, {Owen}, {Pain}, {Palecek}, {Parejko},
  {Parsons}, {Pease}, {Peterson}, {Peterson}, {Petravick}, {Libby Petrick},
  {Petry}, {Pierfederici}, {Pietrowicz}, {Pike}, {Pinto}, {Plante}, {Plate},
  {Plutchak}, {Price}, {Prouza}, {Radeka}, {Rajagopal}, {Rasmussen},
  {Regnault}, {Reil}, {Reiss}, {Reuter}, {Ridgway}, {Riot}, {Ritz}, {Robinson},
  {Roby}, {Roodman}, {Rosing}, {Roucelle}, {Rumore}, {Russo}, {Saha},
  {Sassolas}, {Schalk}, {Schellart}, {Schindler}, {Schmidt}, {Schneider},
  {Schneider}, {Schoening}, {Schumacher}, {Schwamb}, {Sebag}, {Selvy},
  {Sembroski}, {Seppala}, {Serio}, {Serrano}, {Shaw}, {Shipsey}, {Sick},
  {Silvestri}, {Slater}, {Smith}, {Smith}, {Sobhani}, {Soldahl},
  {Storrie-Lombardi}, {Stover}, {Strauss}, {Street}, {Stubbs}, {Sullivan},
  {Sweeney}, {Swinbank}, {Szalay}, {Takacs}, {Tether}, {Thaler}, {Thayer},
  {Thomas}, {Thornton}, {Thukral}, {Tice}, {Trilling}, {Turri}, {Van Berg},
  {Vanden Berk}, {Vetter}, {Virieux}, {Vucina}, {Wahl}, {Walkowicz}, {Walsh},
  {Walter}, {Wang}, {Wang}, {Warner}, {Wiecha}, {Willman}, {Winters},
  {Wittman}, {Wolff}, {Wood-Vasey}, {Wu}, {Xin}, {Yoachim}, \&
  {Zhan}}]{LSST2019}
{Ivezi{\'c}}, {\v{Z}}., {Kahn}, S.~M., {Tyson}, J.~A., {et~al.} 2019, \apj,
  873, 111

\bibitem[{{Joye} \& {Mandel}(2003)}]{ds9}
{Joye}, W.~A. \& {Mandel}, E. 2003, in Astronomical Society of the Pacific
  Conference Series, Vol. 295, Astronomical Data Analysis Software and Systems
  XII, ed. H.~E. {Payne}, R.~I. {Jedrzejewski}, \& R.~N. {Hook}, 489

\bibitem[{{Jullo} \& {Kneib}(2009)}]{jullo2009}
{Jullo}, E. \& {Kneib}, J.~P. 2009, \mnras, 395, 1319

\bibitem[{{Jullo} {et~al.}(2007){Jullo}, {Kneib}, {Limousin},
  {El{\'\i}asd{\'o}ttir}, {Marshall}, \& {Verdugo}}]{jullo2007}
{Jullo}, E., {Kneib}, J.~P., {Limousin}, M., {et~al.} 2007, New Journal of
  Physics, 9, 447

\bibitem[{{Kelly} {et~al.}(2015){Kelly}, {Rodney}, {Treu}, {Foley}, {Brammer},
  {Schmidt}, {Zitrin}, {Sonnenfeld}, {Strolger}, {Graur}, {Filippenko}, {Jha},
  {Riess}, {Bradac}, {Weiner}, {Scolnic}, {Malkan}, {von der Linden}, {Trenti},
  {Hjorth}, {Gavazzi}, {Fontana}, {Merten}, {McCully}, {Jones}, {Postman},
  {Dressler}, {Patel}, {Cenko}, {Graham}, \& {Tucker}}]{kelly2015}
{Kelly}, P.~L., {Rodney}, S.~A., {Treu}, T., {et~al.} 2015, Science, 347, 1123

\bibitem[{{Klein} {et~al.}(2024){Klein}, {Mohr}, {Bocquet}, {Aguena}, {Allen},
  {Alves}, {Ansarinejad}, {Ashby}, {Bacon}, {Bayliss}, {Benson}, {Bleem},
  {Brodwin}, {Brooks}, {Bulbul}, {Burke}, {Canning}, {Carlstrom}, {Rosell},
  {Carretero}, {Chang}, {Conselice}, {Costanzi}, {Crites}, {da Costa},
  {Pereira}, {Davis}, {De Vicente}, {Desai}, {de Haan}, {Dobbs}, {Doel},
  {Ferrero}, {Flores}, {Frieman}, {George}, {Giannini}, {Gladders}, {Gonzalez},
  {Grandis}, {Gruen}, {Gruendl}, {Gutierrez}, {Halverson}, {Hinton}, {Holder},
  {Hollowood}, {Holzapfel}, {Honscheid}, {Hrubes}, {Huang}, {James}, {Khullar},
  {Kim}, {Knox}, {Kraft}, {K{\'e}ruzor{\'e}}, {Lee}, {Luong-Van}, {Mahler},
  {Mantz}, {Marrone}, {Marshall}, {McDonald}, {McMahon}, {Mena-Fern{\'a}ndez},
  {Menanteau}, {Meyer}, {Miquel}, {Myles}, {Padin}, {Pieres}, {Plazas
  Malag{\'o}n}, {Pryke}, {Reichardt}, {Reil}, {Roberson}, {Romer}, {Romero},
  {Ruhl}, {Saliwanchik}, {Salvati}, {Sanchez}, {Saro}, {Schaffer},
  {Schrabback}, {Schubnell}, {Sevilla-Noarbe}, {Sharon}, {Shirokoff}, {Smith},
  {Somboonpanyakul}, {Stalder}, {Stanford}, {Stark}, {Strazzullo}, {Suchyta},
  {Swanson}, {Tarle}, {To}, {Vanderlinde}, {Vieira}, {von der Linden},
  {Weaverdyck}, {Williamson}, {Wiseman}, \& {Young}}]{klein2024}
{Klein}, M., {Mohr}, J.~J., {Bocquet}, S., {et~al.} 2024, \mnras, 531, 3973

\bibitem[{{Kneib} {et~al.}(1996){Kneib}, {Ellis}, {Smail}, {Couch}, \&
  {Sharples}}]{kneib1996}
{Kneib}, J.~P., {Ellis}, R.~S., {Smail}, I., {Couch}, W.~J., \& {Sharples},
  R.~M. 1996, \apj, 471, 643

\bibitem[{{Lang}(2014)}]{lang2014}
{Lang}, D. 2014, \aj, 147, 108

\bibitem[{{Lemon} {et~al.}(2022){Lemon}, {Millon}, {Sluse}, {Courbin}, {Auger},
  {Chan}, {Paic}, \& {Agnello}}]{lemon2022}
{Lemon}, C., {Millon}, M., {Sluse}, D., {et~al.} 2022, \aap, 657, A113

\bibitem[{{Liu} {et~al.}(2011){Liu}, {Shen}, {Strauss}, \& {Hao}}]{liu2011}
{Liu}, X., {Shen}, Y., {Strauss}, M.~A., \& {Hao}, L. 2011, \apj, 737, 101

\bibitem[{{Lombardi}(2024)}]{lombardi2024}
{Lombardi}, M. 2024, arXiv e-prints, arXiv:2406.15280

\bibitem[{{Mannucci} {et~al.}(2023){Mannucci}, {Scialpi}, {Ciurlo}, {Yeh},
  {Marconcini}, {Tozzi}, {Cresci}, {Marconi}, {Amiri}, {Belfiore}, {Carniani},
  {Cicone}, {Nardini}, {Pancino}, {Rubinur}, {Severgnini}, {Ulivi}, {Venturi},
  {Vignali}, {Volonteri}, {Pinna}, {Rossi}, {Puglisi}, {Agapito}, {Plantet},
  {Ghose}, {Carbonaro}, {Xompero}, {Grani}, {Esposito}, {Power}, {Guerra
  Ramon}, {Lefebvre}, {Cavallaro}, {Davies}, {Riccardi}, {Macintosh}, {Taylor},
  {Dolci}, {Baruffolo}, {Feuchtgruber}, {Kravchenko}, {Rau}, {Sturm},
  {Wiezorrek}, {Dallilar}, \& {Kenworthy}}]{mannucci2023}
{Mannucci}, F., {Scialpi}, M., {Ciurlo}, A., {et~al.} 2023, \aap, 680, A53

\bibitem[{{Martinez} {et~al.}(2023){Martinez}, {Napier}, {Cloonan}, {Sukay},
  {Gozman}, {Merz}, {Khullar}, {Lin}, {Matthews Acu{\~n}a}, {Medina},
  {Sanchez}, {Sisco}, {Kavin Stein}, {Tavangar}, {Gonz{\'a}lez}, {Mahler},
  {Sharon}, {Dahle}, \& {Gladders}}]{martinez2023}
{Martinez}, M.~N., {Napier}, K.~A., {Cloonan}, A.~P., {et~al.} 2023, \apj, 946,
  63

\bibitem[{{Moresco} {et~al.}(2022){Moresco}, {Amati}, {Amendola}, {Birrer},
  {Blakeslee}, {Cantiello}, {Cimatti}, {Darling}, {Della Valle}, {Fishbach},
  {Grillo}, {Hamaus}, {Holz}, {Izzo}, {Jimenez}, {Lusso}, {Meneghetti},
  {Piedipalumbo}, {Pisani}, {Pourtsidou}, {Pozzetti}, {Quartin}, {Risaliti},
  {Rosati}, \& {Verde}}]{moresco2022}
{Moresco}, M., {Amati}, L., {Amendola}, L., {et~al.} 2022, Living Reviews in
  Relativity, 25, 6

\bibitem[{{Napier} {et~al.}(2023){Napier}, {Gladders}, {Sharon}, {Dahle},
  {Cloonan}, {Mahler}, {Escapa}, {Garza}, {Kisare}, {Malagon}, {Mork}, {Niu},
  {Rosener}, {Sullivan}, {Tagliavia}, {Tamargo-Arizmendi}, {Teixeira},
  {Tsiane}, {Wagner}, {Zhang}, \& {Zhao}}]{napier2023}
{Napier}, K., {Gladders}, M.~D., {Sharon}, K., {et~al.} 2023, \apjl, 954, L38

\bibitem[{{Predehl} {et~al.}(2021){Predehl}, {Andritschke}, {Arefiev},
  {Babyshkin}, {Batanov}, {Becker}, {B{\"o}hringer}, {Bogomolov}, {Boller},
  {Borm}, {Bornemann}, {Br{\"a}uninger}, {Br{\"u}ggen}, {Brunner}, {Brusa},
  {Bulbul}, {Buntov}, {Burwitz}, {Burkert}, {Clerc}, {Churazov}, {Coutinho},
  {Dauser}, {Dennerl}, {Doroshenko}, {Eder}, {Emberger}, {Eraerds},
  {Finoguenov}, {Freyberg}, {Friedrich}, {Friedrich}, {F{\"u}rmetz},
  {Georgakakis}, {Gilfanov}, {Granato}, {Grossberger}, {Gueguen}, {Gureev},
  {Haberl}, {H{\"a}lker}, {Hartner}, {Hasinger}, {Huber}, {Ji}, {Kienlin},
  {Kink}, {Korotkov}, {Kreykenbohm}, {Lamer}, {Lomakin}, {Lapshov}, {Liu},
  {Maitra}, {Meidinger}, {Menz}, {Merloni}, {Mernik}, {Mican}, {Mohr},
  {M{\"u}ller}, {Nandra}, {Nazarov}, {Pacaud}, {Pavlinsky}, {Perinati},
  {Pfeffermann}, {Pietschner}, {Ramos-Ceja}, {Rau}, {Reiffers}, {Reiprich},
  {Robrade}, {Salvato}, {Sanders}, {Santangelo}, {Sasaki}, {Scheuerle},
  {Schmid}, {Schmitt}, {Schwope}, {Shirshakov}, {Steinmetz}, {Stewart},
  {Str{\"u}der}, {Sunyaev}, {Tenzer}, {Tiedemann}, {Tr{\"u}mper}, {Voron},
  {Weber}, {Wilms}, \& {Yaroshenko}}]{predehl2021}
{Predehl}, P., {Andritschke}, R., {Arefiev}, V., {et~al.} 2021, \aap, 647, A1

\bibitem[{{Price-Whelan} {et~al.}(2018){Price-Whelan}, {Sip{\H{o}}cz},
  {G{\"u}nther}, {Lim}, {Crawford}, {Conseil}, {Shupe}, {Craig}, {Dencheva},
  {Ginsburg}, {VanderPlas}, {Bradley}, {P{\'e}rez-Su{\'a}rez}, {de Val-Borro},
  {Paper Contributors}, {Aldcroft}, {Cruz}, {Robitaille}, {Tollerud},
  {Coordination Committee}, {Ardelean}, {Babej}, {Bach}, {Bachetti}, {Bakanov},
  {Bamford}, {Barentsen}, {Barmby}, {Baumbach}, {Berry}, {Biscani}, {Boquien},
  {Bostroem}, {Bouma}, {Brammer}, {Bray}, {Breytenbach}, {Buddelmeijer},
  {Burke}, {Calderone}, {Cano Rodr{\'\i}guez}, {Cara}, {Cardoso}, {Cheedella},
  {Copin}, {Corrales}, {Crichton}, {D{\textquoteright}Avella}, {Deil},
  {Depagne}, {Dietrich}, {Donath}, {Droettboom}, {Earl}, {Erben}, {Fabbro},
  {Ferreira}, {Finethy}, {Fox}, {Garrison}, {Gibbons}, {Goldstein}, {Gommers},
  {Greco}, {Greenfield}, {Groener}, {Grollier}, {Hagen}, {Hirst}, {Homeier},
  {Horton}, {Hosseinzadeh}, {Hu}, {Hunkeler}, {Ivezi{\'c}}, {Jain}, {Jenness},
  {Kanarek}, {Kendrew}, {Kern}, {Kerzendorf}, {Khvalko}, {King}, {Kirkby},
  {Kulkarni}, {Kumar}, {Lee}, {Lenz}, {Littlefair}, {Ma}, {Macleod},
  {Mastropietro}, {McCully}, {Montagnac}, {Morris}, {Mueller}, {Mumford},
  {Muna}, {Murphy}, {Nelson}, {Nguyen}, {Ninan}, {N{\"o}the}, {Ogaz}, {Oh},
  {Parejko}, {Parley}, {Pascual}, {Patil}, {Patil}, {Plunkett}, {Prochaska},
  {Rastogi}, {Reddy Janga}, {Sabater}, {Sakurikar}, {Seifert}, {Sherbert},
  {Sherwood-Taylor}, {Shih}, {Sick}, {Silbiger}, {Singanamalla}, {Singer},
  {Sladen}, {Sooley}, {Sornarajah}, {Streicher}, {Teuben}, {Thomas},
  {Tremblay}, {Turner}, {Terr{\'o}n}, {van Kerkwijk}, {de la Vega}, {Watkins},
  {Weaver}, {Whitmore}, {Woillez}, {Zabalza}, \& {Contributors}}]{astropy2}
{Price-Whelan}, A.~M., {Sip{\H{o}}cz}, B.~M., {G{\"u}nther}, H.~M., {et~al.}
  2018, \aj, 156, 123

\bibitem[{{Prochaska} {et~al.}(2020){Prochaska}, {Hennawi}, {Westfall},
  {Cooke}, {Wang}, {Hsyu}, {Davies}, {Farina}, \& {Pelliccia}}]{prochaska20}
{Prochaska}, J., {Hennawi}, J., {Westfall}, K., {et~al.} 2020, The Journal of
  Open Source Software, 5, 2308

\bibitem[{{Refsdal}(1964)}]{refsdal1964}
{Refsdal}, S. 1964, \mnras, 128, 307

\bibitem[{Rohatgi(2024)}]{rohatgi2024}
Rohatgi, A. 2024, WebPlotDigitizer

\bibitem[{{Schlafly} {et~al.}(2019){Schlafly}, {Meisner}, \&
  {Green}}]{schlafly2019}
{Schlafly}, E.~F., {Meisner}, A.~M., \& {Green}, G.~M. 2019, \apjs, 240, 30

\bibitem[{{Shu} {et~al.}(2019){Shu}, {Koposov}, {Evans}, {Belokurov},
  {McMahon}, {Auger}, \& {Lemon}}]{shu2019}
{Shu}, Y., {Koposov}, S.~E., {Evans}, N.~W., {et~al.} 2019, \mnras, 489, 4741

\bibitem[{{Shu} {et~al.}(2018){Shu}, {Marques-Chaves}, {Evans}, \&
  {P{\'e}rez-Fournon}}]{shu2018}
{Shu}, Y., {Marques-Chaves}, R., {Evans}, N.~W., \& {P{\'e}rez-Fournon}, I.
  2018, \mnras, 481, L136

\bibitem[{{Sluse} {et~al.}(2012){Sluse}, {Hutsem{\'e}kers}, {Courbin},
  {Meylan}, \& {Wambsganss}}]{sluse2012}
{Sluse}, D., {Hutsem{\'e}kers}, D., {Courbin}, F., {Meylan}, G., \&
  {Wambsganss}, J. 2012, \aap, 544, A62

\bibitem[{{Suyu} {et~al.}(2017){Suyu}, {Bonvin}, {Courbin}, {Fassnacht},
  {Rusu}, {Sluse}, {Treu}, {Wong}, {Auger}, {Ding}, {Hilbert}, {Marshall},
  {Rumbaugh}, {Sonnenfeld}, {Tewes}, {Tihhonova}, {Agnello}, {Blandford},
  {Chen}, {Collett}, {Koopmans}, {Liao}, {Meylan}, \& {Spiniello}}]{suyu2017}
{Suyu}, S.~H., {Bonvin}, V., {Courbin}, F., {et~al.} 2017, \mnras, 468, 2590

\bibitem[{{Taylor}(2005)}]{topcat}
{Taylor}, M.~B. 2005, in Astronomical Society of the Pacific Conference Series,
  Vol. 347, Astronomical Data Analysis Software and Systems XIV, ed.
  P.~{Shopbell}, M.~{Britton}, \& R.~{Ebert}, 29

\bibitem[{{Treu} {et~al.}(2022){Treu}, {Suyu}, \& {Marshall}}]{treu2022}
{Treu}, T., {Suyu}, S.~H., \& {Marshall}, P.~J. 2022, \aapr, 30, 8

\bibitem[{{van der Walt} {et~al.}(2011){van der Walt}, {Colbert}, \&
  {Varoquaux}}]{numpy1}
{van der Walt}, S., {Colbert}, S.~C., \& {Varoquaux}, G. 2011, Computing in
  Science Engineering, 13, 22

\bibitem[{Van~Rossum \& Drake(2009)}]{python}
Van~Rossum, G. \& Drake, F.~L. 2009, Python 3 Reference Manual (Scotts Valley,
  CA: CreateSpace)

\bibitem[{{Verde} {et~al.}(2019){Verde}, {Treu}, \& {Riess}}]{verde2019}
{Verde}, L., {Treu}, T., \& {Riess}, A.~G. 2019, Nature Astronomy, 3, 891

\bibitem[{{Virtanen} {et~al.}(2020){Virtanen}, {Gommers}, {Oliphant},
  {Haberland}, {Reddy}, {Cournapeau}, {Burovski}, {Peterson}, {Weckesser},
  {Bright}, {van der Walt}, {Brett}, {Wilson}, {Jarrod Millman}, {Mayorov},
  {Nelson}, {Jones}, {Kern}, {Larson}, {Carey}, {Polat}, {Feng}, {Moore}, {Vand
  erPlas}, {Laxalde}, {Perktold}, {Cimrman}, {Henriksen}, {Quintero}, {Harris},
  {Archibald}, {Ribeiro}, {Pedregosa}, {van Mulbregt}, \&
  {Contributors}}]{scipy}
{Virtanen}, P., {Gommers}, R., {Oliphant}, T.~E., {et~al.} 2020, Nature
  Methods, 17, 261

\bibitem[{{Wen} \& {Han}(2022)}]{wen2022}
{Wen}, Z.~L. \& {Han}, J.~L. 2022, \mnras, 513, 3946

\bibitem[{{Wen} \& {Han}(2024)}]{wen2024}
{Wen}, Z.~L. \& {Han}, J.~L. 2024, \apjs, 272, 39

\bibitem[{{Wong} {et~al.}(2020){Wong}, {Suyu}, {Chen}, {Rusu}, {Millon},
  {Sluse}, {Bonvin}, {Fassnacht}, {Taubenberger}, {Auger}, {Birrer}, {Chan},
  {Courbin}, {Hilbert}, {Tihhonova}, {Treu}, {Agnello}, {Ding}, {Jee},
  {Komatsu}, {Shajib}, {Sonnenfeld}, {Blandford}, {Koopmans}, {Marshall}, \&
  {Meylan}}]{wong2020}
{Wong}, K.~C., {Suyu}, S.~H., {Chen}, G. C.~F., {et~al.} 2020, \mnras, 498,
  1420

\bibitem[{{Wright} {et~al.}(2010){Wright}, {Eisenhardt}, {Mainzer}, {Ressler},
  {Cutri}, {Jarrett}, {Kirkpatrick}, {Padgett}, {McMillan}, {Skrutskie},
  {Stanford}, {Cohen}, {Walker}, {Mather}, {Leisawitz}, {Gautier}, {McLean},
  {Benford}, {Lonsdale}, {Blain}, {Mendez}, {Irace}, {Duval}, {Liu}, {Royer},
  {Heinrichsen}, {Howard}, {Shannon}, {Kendall}, {Walsh}, {Larsen}, {Cardon},
  {Schick}, {Schwalm}, {Abid}, {Fabinsky}, {Naes}, \& {Tsai}}]{wise2010}
{Wright}, E.~L., {Eisenhardt}, P. R.~M., {Mainzer}, A.~K., {et~al.} 2010, \aj,
  140, 1868

\end{thebibliography}

\end{document}